\documentclass[twocolumn,prb,amsmath,amssymb,floatfix,superscriptaddress,showpacs]{revtex4}
\usepackage{graphicx}
\usepackage{dcolumn}
\usepackage{bm}
\usepackage{times}

\begin{document}

\title{Robust charge and magnetic order under electric field/current
in the multiferroic LuFe$_2$O$_4$}

\author{Jinsheng Wen}
\affiliation{Condensed Matter Physics and Materials Science
Department, Brookhaven National Laboratory, Upton, New York 11973,
USA}
\affiliation{Department of Materials Science and Engineering,
Stony Brook University, Stony Brook, New York 11794, USA}
\author{Guangyong Xu}
\affiliation{Condensed Matter Physics and Materials Science
Department, Brookhaven National Laboratory, Upton, New York 11973,
USA}
\author{Genda Gu}
\affiliation{Condensed Matter Physics and Materials Science
Department, Brookhaven National Laboratory, Upton, New York 11973,
USA}
\author{S.~M.~Shapiro}
\affiliation{Condensed Matter Physics and Materials Science
Department, Brookhaven National Laboratory, Upton, New York 11973,
USA}
\date{\today}

\begin{abstract}
We performed elastic neutron scattering measurements on the charge-
and magnetically-ordered multiferroic material LuFe$_2$O$_4$. An
external electric field along the [001] direction with strength up
to 20~kV/cm applied at low temperature ($\sim$~100~K) does not
affect either the charge or magnetic structure. At higher
temperatures ($\sim$~360~K), before the transition to
three-dimensional charge-ordered state, the resistivity of the
sample is low, and an electric current was applied instead. A
reduction of the charge and magnetic peak intensities occurs when
the sample is cooled under a constant electric current. However,
after calibrating the real sample temperature using its own
resistance-temperature curve, we show that the actual sample
temperature is higher than the thermometer readings, and the
``intensity reduction'' is entirely due to internal sample heating
by the applied current. Our results suggest that the charge and
magnetic orders in LuFe$_2$O$_4$ are unaffected by the application
of external electric field/current, and previously observed electric
field/current effects can be naturally explained by internal sample
heating.
\end{abstract}


\maketitle

LuFe$_2$O$_4$ is a new multiferroic material. Here the bulk
ferroelectric polarization is not due to cation displacements as in
conventional ferroelectrics, but instead arises from the
three-dimensional charge valence order of Fe$^{2+}$ and Fe$^{3+}$
ions occurring at $\sim$~340~K.~\cite{ikeda} The magnetic order
starts at lower temperature ($\sim$~240~K) in the charge-ordered
ferroelectric phase.~\cite{wen} Recent reports of strong couplings
between the two orders~\cite{ikeda,angst-2008-101}, as well as large
room temperature magneto-electric response in this
material~\cite{Subramanian,wen}, make LuFe$_2$O$_4$ a promising
candidate for practical applications.

In addition to the magneto-electric response, tremendous interest
has been focused on studying the electric-field response of the
magnetic structures in multiferroic systems. Nevertheless, there
have been only a few observations of such
effects~\cite{efield_1,efield_2,econtrol1,efield_3,efieldp1,efieldp2}
in known multiferroic systems, and all of them can be attributed to
electric field realigning ferroelectric domains and therefore
causing a macroscopic magnetic response. In LuFe$_2$O$_4$, the
ferroelectric polarization is charge-valence driven, and the
charge-valence order also couples strongly to the magnetic order. If
the charge order can be affected or broken, it is then possible to
affect the microscopic magnetic structures by an external electric
field or current. Indeed, there have been previous reports on
non-linear current-voltage behaviors, and eventually an
electric-field ``breakdown'' of the charge order in
LuFe$_2$O$_4$~\cite{li-2008-93,zeng-2008}. There have also been
claims of electrical control of the magnetic response in the same
material.~\cite{li:172412} These observations, if confirmed and
fully understood, would be extremely interesting and important for
achieving mutual control of electric and magnetic degrees of
freedoms in multiferroic systems.

We thus performed elastic neutron scattering measurements on single
crystals of LuFe$_2$O$_4$, studying the response of the charge- and
magnetic-order Bragg peaks under external electric field/current. No
electric-field effect has been observed at low temperature
$\sim$~100~K for a field strength up to 20~kV/cm. Near room
temperature an electric current effect on the ordering is observed,
and we show that it is due to internal heating of the sample by the
current flowing through the sample. We conclude that the charge and
magnetic order are robust and not affected by the electric
field/current.

\begin{figure}[ht]
\includegraphics[width=\linewidth]{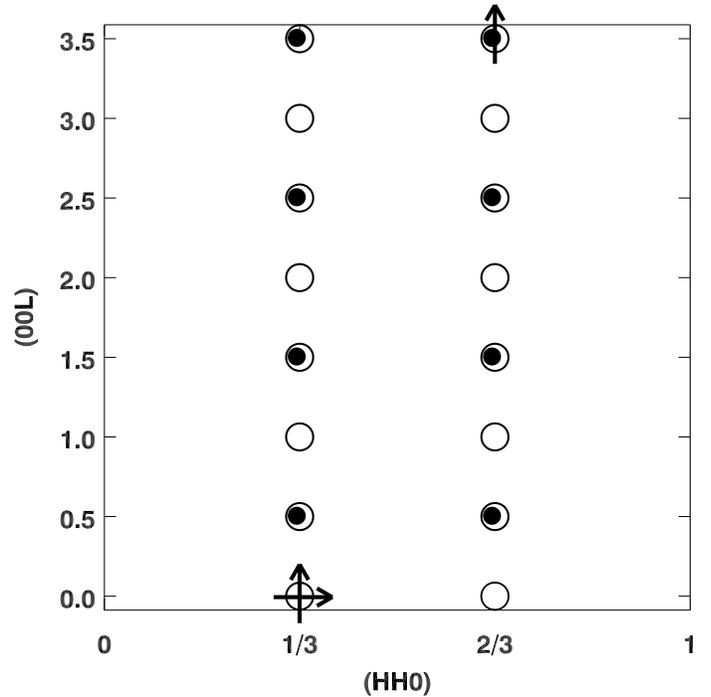}
\caption{Schematic of the magnetic- (open circles) and charge-order
(filled circles) peaks in reciprocal space. The arrows indicate the
scan directions along [110] and [001].} \label{fig:1}
\end{figure}

Single crystals LuFe$_2$O$_4$ were grown using floating zone
technique.~\cite{crystalgrowth} Typical crystal sizes are
$\sim$~$10\times5\times2$~mm$^3$. Our neutron scattering
measurements were performed on BT9 triple-axis-spectrometer at the
NIST Center for Neutron Research. An incident neutron energy of
14.7~meV was selected by a pyrographic (PG002) monochromator, with
beam collimations of $40'$-$40'$-$40'$-$80'$, and another PG002
crystal was used as the analyzer. PG filters were used before the
sample to reduce background from higher order neutrons. A sample of
0.9~g was loaded in a closed-cycle refrigerator, where the
thermometer is attached to the mounting base, about 3~cm away from
the sample. LuFe$_2$O$_4$ has a hexagonal structure with three iron
double layers in each unit cell. The ferroelectric polarization is
directly due to the imbalance of iron valences in each double layer,
and the net (induced) polarization appears along the [001]
direction, perpendicular to the hexagonal plane and the double
layers. The two $10\times5$~mm$^2$ (001) surfaces were painted with
silver paint so that electric field/current can be applied along the
[001] direction (2~mm thick) for our measurements. The single
crystal sample was oriented so that the horizontal diffraction plane
is the $(HHL)$ plane, defined by the vectors [110] and [001]. (See
Fig.~\ref{fig:1}). The resistance-temperature (R-T) curve has been
measured using Keitheley 2000 multimeter, and the in-situ resistance
during the neutron scattering measurements was obtained by reading
the voltage across the sample while keeping the current constant.

The magnetic Bragg peaks in this compound occur at reciprocal space
positions (1/3, 1/3, $L$) and (2/3, 2/3, $L$) for both half-integer
and integer $L$ values~\cite{twodspin,christianson:107601} below the
magnetic ordering temperature $T_N \sim$~240~K; while the charge
peaks only appear at half-integer $L$
values~\cite{PhysRevB.62.12167,
zhang:247602,angst-2008-101,wen,christianson:107601}. (See
Fig.~\ref{fig:1}).

\begin{figure}[ht]
\includegraphics[height=0.9\linewidth,angle=90]{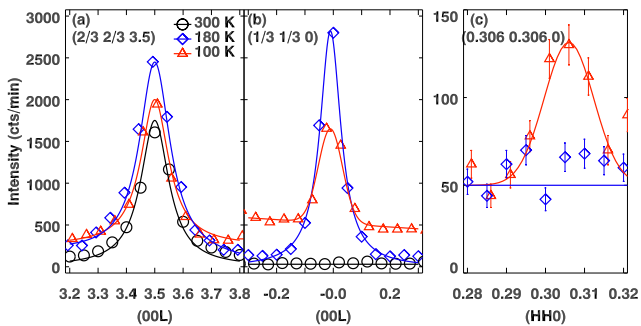}
\caption{(Color Online) Scans through (a) charge-order peak (2/3,
2/3, 3.5) along [001] direction at different temperatures; (b)
magnetic peak (1/3, 1/3, 0) along [001] direction; (c) satellite
peak (0.306, 0.306, 0) along [110] direction. Error bars represent
square root of the total counts, and those in (a) and (b) are
smaller than the symbols. Lines through data are guides to the
eyes.} \label{fig:2}
\end{figure}

\begin{figure}[ht]
\includegraphics[width=0.9\linewidth]{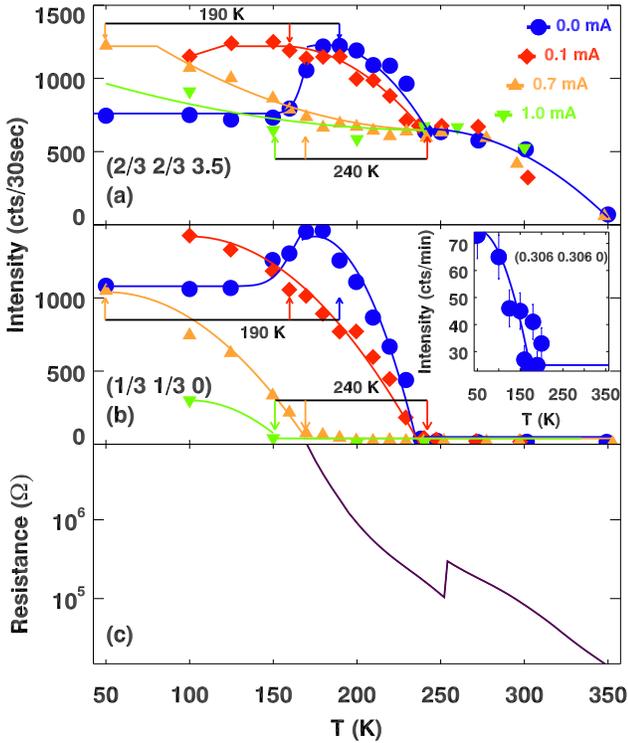}
\caption{(Color Online) (a) Charge, and (b) magnetic peak intensity
as a function of thermometer temperature, obtained under cooling
with different currents. Inset in (b) shows temperature dependence
of the satellite peak (0.306, 0.306, 0) intensity. Error bars
represent square root of the total counts, and those smaller than
the symbols are not shown. Lines through data are guides to the
eyes. Arrows indicate the actual temperatures determined by
measuring the resistance. (c) R-T curve measured with nearly zero
($\sim~1~\mu$A) current. The resistance below 170~K exceeds the
maximum of the multimeter.} \label{fig:3}
\end{figure}

We choose to monitor (2/3, 2/3, 3.5) for the charge order, and (1/3,
1/3, 0) for the magnetic order. Representative scans through these
two peaks are plotted in Fig.~\ref{fig:2}(a) and (b) respectively.
These peak intensities indicate the charge/magnetic order and are
plotted vs. temperature in Fig.~\ref{fig:3}(a) and (b). For zero
field, the charge order starts as two-dimensional (2D) at around
500~K~\cite{PhysRevB.62.12167}, and becomes three-dimensional (3D)
around $T_{CO}\sim$~340~K~\cite{wen,christianson:107601}. At
$T_N\sim$~240~K, the magnetic order occurs, as indicated by the rise
of (1/3,1/3,0) intensity shown in Fig.~\ref{fig:3}(b). Also, a boost
to the intensity at (2/3,2/3, 3.5) is observed [see
Fig.~\ref{fig:3}(a)], which is due to the additional scattering
intensity at this wave-vector coming from magnetic ordering. The
intensity at (2/3,2/3,3.5) now (for $T<T_N$) has contributions from
both the charge and magnetic orders. With further cooling, the
intensity increases until reaching another temperature
$T_L\sim$~180~K. Here another phase transition occurs, similar to
that observed in Ref.~\onlinecite{christianson:107601}. Note that
this second phase transition is strongly sample dependent, and has
been shown to be missing for some
samples.~\cite{wu:137203,PhysRevB.62.12167,wen} As demonstrated by
previous studies, the magnetic properties of LuFeO$_{4+\delta}$ are
sensitive to oxygen stoichiometry,~\cite{wang:024419} and different
temperature dependence of intensity below $T_L \sim$~180~K is likely
due to different oxygen content in different samples. The
intensities for both peaks drop at $T_L$ but become almost flat
below $\sim$~150~K [Fig.~\ref{fig:3}(a) and (b)]. This is also
demonstrated in Fig.~\ref{fig:2}, where at 100~K both the charge and
magnetic peak intensities are lower than those at 180~K.
Additionally, at 100~K there is a strong 2D diffuse type magnetic
scattering~\cite{christianson:107601} which shows up as higher
``background'' in the $L$-scans [Fig.~\ref{fig:2}(b)]. In addition,
intensity starts to appear at satellite positions below $T_L$, and
in Fig.~\ref{fig:2}(c) we plot scans through a satellite peak around
(0.306, 0.306, 0), whose intensity dependence on the temperature is
plotted in the inset of Fig.~\ref{fig:3}(b).

These results suggest that the low-temperature (magnetic/charge)
structures of LuFe$_2$O$_4$ are quite complicated and sometimes
sample dependent. Nevertheless, our goal is to search for possible
electric-field effect on the magnetic and charger orders. As
suggested by Angst {\it et al.}, the energy difference between
antiferroelectric and ferroelectric charge-order configuration is
only $\sim$~3\%, and it is possible to stabilize the ferroelectric
configuration when the system is cooled in an electric
field.~\cite{angst-2008-101} The low resistivity
($\sim$~$10^3~\Omega\cdot$m at room temperature) makes it very
difficult to apply a static electric field and do the field-cooling
measurements. Instead, we applied an electric field of 20~kV/cm
along the [001] direction at 100~K, and performed scans through
(2/3, 2/3, 3.5), (1/3, 1/3, 0), and (0.306, 0.306, 0) peaks. The
scan profiles are identical to those shown in Fig.~\ref{fig:2},
which indicates that there is no observable electric-field effect
with fields applied below $T_{CO}$. At 100~K, the current is
estimated to be on the order of $\mu$A.

\begin{figure}[ht]
\includegraphics[height=0.9\linewidth,angle=90]{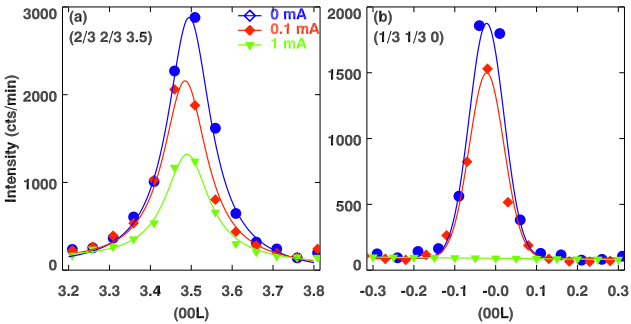}
\caption{(Color Online) (a) Charge (2/3, 2/3, 3.5), and (b) magnetic
(1/3, 1/3, 0) peak measured at 200~K, after cooled with different
currents applied at 360~K. Error bars are smaller than the symbols.
Lines through data are guides to the eyes.} \label{fig:4}
\end{figure}

To investigate the response of the system to electric currents, we
applied different currents at 360~K, cooled the system with the
current maintained as constant, and performed scans at different
temperatures. Results of the scans through the charge and magnetic
peaks at 200~K with different currents applied are shown in
Fig.~\ref{fig:4}. It is clear that for both charge and magnetic
peaks, the peak intensities are reduced when cooled under electric
current, and the intensity reduction increases with increasing
current. The magnetic peak at 200~K is fully suppressed by a 1~mA
current. To further examine the current effect, we plotted the
charge and magnetic order peak intensity as a function of
thermometer temperature with different currents in
Fig.~\ref{fig:3}(a) and (b).

\begin{figure}[ht]
\includegraphics[width=0.9\linewidth]{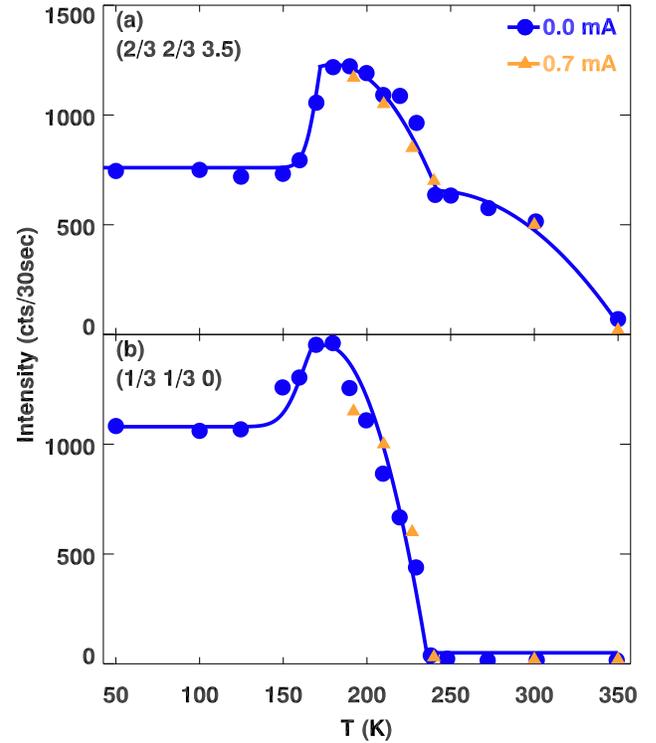}
\caption{(Color Online) (a) Charge, and (b) magnetic peak intensity
as a function of calibrated temperature, obtained under cooling with
a  zero and 0.7~mA current. Error bars are smaller than the symbols.
Lines through data are guides to the eyes.} \label{fig:5}
\end{figure}

When the sample was cooled with current, both charge and magnetic
peak intensities were reduced. The current effects on the intensity
clearly become more pronounced as the current increases. These
results correlate with earlier studies on the electric field/current
effects on transport and magnetic properties of
LuFe$_2$O$_4$,~\cite{li:182903,li:172412,li-2008-93,zeng-2008} and
the current reducing the peak intensity here can be attributed to
the current inducing breakdown of the charge order, as suggested in
Refs.~\onlinecite{li-2008-93,zeng-2008}. However, there are also
indications that this may not be as simple. One indication is that
with 1~mA current cooling, the temperature reading never went below
100~K. This suggests that there is a significant heating power
applied to the sample (by the current applied). Another observation
is that if we remove the current at 100~K, the peak intensity does
not return to the ZFC value immediately. Instead, there is a 30 to
60 seconds lag for the (charge peak) intensity to fully recover with
a current of 1~mA , while at the same time, the temperature reading
is constant. This time scale is too long for any real charge diffuse
process to occur in these materials, and is a strong evidence that
internal sample heating is playing a role. Because the thermometer
is attached to the base of the sample mounting post, which is about
3~cm away from the sample position, it is plausible that there could
be a large temperature gradient between the sample position and the
thermometer location.

In order to calibrate the sample temperature, we use the temperature
dependence of resistance of the sample along $c$-direction as an
independent measure as shown in Fig.~\ref{fig:3}(c). The current
used to measure the resistance is small, on the order of $1~\mu$A,
so that the condition under which the resistance was measured can be
taken as zero-current cooling. Under cooling, the resistance
increases continuously through $T_{CO}$, which is consistent with
the observation that below $T_{CO}$, charge order is still
short-ranged along $c$-axis, and disorder is playing an important
role in the material's properties.~\cite{wen} Around 250~K, the
resistance drops, which is related to the magnetic phase transition,
also suggesting a strong coupling between the magnetic and
electrical properties in this material. Below 250~K, the resistance
increases monotonically again.

The R-T curve provides us a good measure of the instantaneous sample
temperature. During the neutron diffraction measurements, we
measured the sample's resistance by reading out the voltage across
the sample while maintaining a constant current when cooling.
Comparing the measured resistance with the R-T curve, we found that
the actual sample temperatures under current-cooling are higher than
those read by the thermometer. The horizontal lines indicate the
``real'' sample temperature (240~K or 190~K) determined by the R-T
curve, and the arrows indicate the temperatures read by the
thermometer. For a sample temperature of 240~K, the thermometer
reads 240~K, 170~K, or 150~K with a current of 0.1~mA, 0.7~mA, or
1~mA respectively. If we correct the temperature scales of the
various cooling curves using the real sample temperatures based on
resistance readings, there is no field effect on the ordering. This
is demonstrated in Fig.~\ref{fig:3}(a) and (b). On each cooling
curve (with different current applied), the data points where the
real sample temperatures are 240~K (or 190~K) are marked by arrows
of different colors, and they indeed have the same magnetic/charge
peak intensities independent of cooling conditions. In
Fig.~\ref{fig:5}, we plot the charge and magnetic peak intensity as
a function of calibrated temperature for cooling with a 0.7~mA
current, and compare them with those under zero-current cooling. It
is clear that there is no real effect if we take out the internal
heating effect. The current-heating effect is not entirely
unexpected, since the specific heat of LuFe$_2$O$_4$ is relatively
low, $\sim$~0.5~J/(K$\cdot$g) at 300~K.~\cite{wang:024419} With a
small sample ($<1$~g), the heater with a power on the order of
$0.1$~W, which corresponds to a current of 1~mA through the sample
at 300~K, or 0.4~mA at 200~K, is high enough to effectively heat the
sample.

This naturally explains the unusual behaviors in LuFe$_2$O$_4$
observed by other groups.~\cite{li:172412,li-2008-93,zeng-2008} The
resistivity is low ($\sim~10^3~\Omega\cdot$m) in the temperature
range where most measurements are carried out, so a small voltage is
able to drive a large current through the sample and heat the sample
significantly. The actual sample temperature in these cases will be
higher than the readings from the thermometer. The material's
magnetization,~\cite{li:172412} and transition
temperature~\cite{li-2008-93,zeng-2008} will then appear to be
affected by the field. The internal current heating can also explain
the observed non-linear current-voltage
behavior.~\cite{li-2008-93,zeng-2008} In addition, because the
resistance decreases with heating---if a constant voltage is applied
to the sample, the sample is heated and the resistance lowers, which
in turn increases the current further, and puts more thermal power
on the sample, which again raises the sample temperature and lowers
its resistance. Eventually an avalanche occurs, which was
interpreted as the ``breakdown'' of the
charge-order.~\cite{li-2008-93,zeng-2008}

Here we observed that the charge order in LuFe$_2$O$_4$ remains
intact with electrical inputs---neither high electric fields applied
at low temperature, nor electric currents applied at high
temperature can affect it. However, the charge order seems to be
rather sensitive to magnetic field, even when no magnetic order is
present.~\cite{wen} The fact that a charge-ordered system is
magnetically sensitive instead of electrically sensitive makes
LuFe$_2$O$_4$ very unusual. In other charge-ordered systems,
electric field is able to slide, or cause breakdown of the charge
order.~\cite{charge_1,charge_2,charge_3} Apparently, from our data,
it is not the case for LuFe$_2$O$_4$. It is likely that in
LuFe$_2$O$_4$ the pinning of the charge order, e.g., by impurities,
is stronger than that in other charge-ordered systems, which makes
it less electrically sensitive.

In summary, we report that the charge and magnetic order in
LuFe$_2$O$_4$ is not affected by electric field (up to 20~kV/cm) or
current. The observed reduction of charge- and magnetic-order peak
intensity is due to resistive heating. Our results also suggest that
electric field/current effects on LuFe$_2$O$_4$, as well as the
non-linear current-voltage behavior reported elsewhere are results
of internal current heating of the sample. This case is very similar
to those observed in charge-stripe ordered cuprates and
nickelates~\cite{break_1,break_2,break_3}, where an electric-field
effect was observed on the charge order, but later entirely
attributed to resistive heating effects~\cite{hucker}.

We thank W. Ratcliff, and M. H\"{u}cker for helpful discussions.
Work at Brookhaven National Laboratory is supported by U.S.
Department of Energy Contract DE-AC02-98CH20886.


\end{document}